\documentstyle[aps,prc,epsfig]{revtex}
\draft

\newcommand{\ba}{\begin{eqnarray}}
\newcommand{\ea}{\end{eqnarray}}
\def\ii{\'{\i}}
\begin{document}
\pagestyle{plain}

\title{Mean-field analysis of interacting boson models 
with random interactions}
\author{R. Bijker$^{a}$ and A. Frank$^{a,b}$}
\address{$^{a}$Instituto de Ciencias Nucleares, 
Universidad Nacional Aut\'onoma de M\'exico, \\
Apartado Postal 70-543, 04510 M\'exico, D.F., M\'exico
\newline
$^{b}$Centro de Ciencias F\ii sicas, 
Universidad Nacional Aut\'onoma de M\'exico, \\
Apartado Postal 139-B, 62251 Cuernavaca, Morelos, M\'exico}

\maketitle

\begin{abstract}
We investigate the origin of the regular features observed in numerical 
studies of the interacting boson model with random interactions, in 
particular the dominance of $L=0$ ground states and the occurrence of 
vibrational and rotational band structures. It is shown that all of these 
properties can be interpreted and explained in terms of a Hartree-Bose 
mean-field analysis, in which different regions of the parameter space 
are associated with geometric shapes. The same conclusions hold for the 
vibron model. 
\end{abstract}


In empirical studies of medium and heavy even-even nuclei very regular 
features have been observed, such as the tripartite classification of 
nuclear structure into seniority, anharmonic vibrator and rotor regions 
\cite{Zamfir}. In each of these three regimes, the energy systematics is 
extremely robust, and the transitions between different regions occur very 
rapidly, typically with the addition or removal of only one or two pairs 
of nucleons. Traditionally, this regular behavior has been interpreted 
as a consequence of particular nucleon-nucleon interactions, such as an
attractive pairing force in semimagic nuclei and an attractive
neutron-proton quadrupole-quadrupole interaction for deformed nuclei. 

It came as a surprise, therefore, that recent studies of even-even nuclei 
in the nuclear shell model \cite{JBD,BFP1,JBDT} and in the interacting boson 
model (IBM) \cite{BF1,BF2,BFP2,KZC} with random interactions (centered 
around zero, i.e. equally likely to be attractive or repulsive) displayed a 
high degree of order. Both models showed a marked statistical preference 
($>60 \,\%$) for ground states with $L=0$, despite the random nature of 
the interactions. In addition, in the shell model evidence was found 
for the occurrence of pairing properties \cite{JBDT}, and in the IBM for 
both vibrational and rotational band structures \cite{BF1,BF2}. 

However, these results were obtained from numerical studies. It is thus 
important to gain a better understanding as to why this happens. 
Can these properties be explained in a more intuitive way? Several 
attempts have been made to understand these surprising phenomena, 
most of them for fermionic systems and especially with regards to the 
dominance of $L=0$ ground states. Among others, we mention a study of the 
time-reversal invariance of random interactions \cite{BFP1,BFP2}, the 
connection with the width of the energy distributions \cite{BFP1,BFP2}, 
the distribution of the lowest energy eigenvalues \cite{BF2}, the effect 
of higher-order interactions \cite{BF2,BFP2}, the geometric chaoticity 
of the angular momentum coupling of individual particles \cite{MVZ}, 
the connection with random polynomials \cite{DK}, 
the correlation between wave functions and energies \cite{Kaplan}, 
the probability distribution of matrix elements \cite{Drozdz}, the 
relation with the diagonal matrix elements of the Hamiltonian \cite{ZA}, 
and average energies and variances \cite{victor}. 

In \cite{MVZ} it was shown that the overlap between the ground state 
wave function obtained in a single-$j$ shell model calculation with 
random interactions with the seniority zero state is small, and that 
the distribution of these overlaps follows closely the predictions for 
chaotic wave functions. This suggests that for fermion systems, at least 
for like-nucleons in a single-$j$ shell, the dominance of $L=0$ ground states 
may be due to the geometric chaoticity of randomly coupled individual spins, 
although the observed oscillations with shell size cannot be explained 
by such a simple scheme. 
On the other hand, in the IBM, besides the $L=0$ ground state dominance, 
a strong preponderance of vibrational and rotational values for energy 
ratios of yrast states was found, as well as a strong correlation with the 
corresponding vibrational and rotational values of the quadrupole 
transitions. This has been interpreted as evidence for the 
existence of vibrational and rotational structure \cite{BF1,BF2}.  
It is the purpose of this Rapid Communication to study the origin 
of the regular features observed in the IBM,  
when the parameters are chosen randomly. 

Collective excitations in nuclei are described in the IBM in 
terms of a system of interacting bosons \cite{IBM}. Its building 
blocks are a quadrupole boson $d^{\dagger}$ with $L^P=2^+$ and a 
scalar boson $s^{\dagger}$ with $L^P=0^+$. The total number of bosons 
$N$ is conserved by the IBM Hamiltonian. We consider the most general 
one- and two-body Hamiltonian
\ba
H &=& \frac{1}{N} \left[ H_1 + \frac{1}{N-1} H_2 \right] ~, 
\ea
with
\ba
H_1 &=& \epsilon_s \, s^{\dagger} \cdot \tilde{s}  
+ \epsilon_d \, d^{\dagger} \cdot \tilde{d} ~, 
\nonumber\\
H_2 &=& \frac{1}{2} \, u_0 \, 
(s^{\dagger} \times s^{\dagger})^{(0)} \cdot 
(\tilde{s} \times \tilde{s})^{(0)} 
+ u_2 \, (s^{\dagger} \times d^{\dagger})^{(2)} \cdot 
(\tilde{d} \times \tilde{s})^{(2)} 
\nonumber\\ 
&&+ \sum_{\lambda=0,2,4} \frac{1}{2} \, c_{\lambda} \, 
(d^{\dagger} \times d^{\dagger})^{(\lambda)} \cdot  
(\tilde{d} \times \tilde{d})^{(\lambda)} 
\nonumber\\
&&+ \frac{1}{2\sqrt{2}} \, v_0 \, \left[ 
  (d^{\dagger} \times d^{\dagger})^{(0)} \cdot 
  (\tilde{s} \times \tilde{s})^{(0)} 
+ H.c. \right]
\nonumber\\
&&+ \frac{1}{2} \, v_2 \, \left[ 
  (d^{\dagger} \times d^{\dagger})^{(2)} \cdot 
  (\tilde{d} \times \tilde{s})^{(2)} 
+ H.c. \right] ~, 
\label{ham}
\ea
where the nine coefficients ($\epsilon_s$, $\epsilon_d$, $u_0$, $u_2$, 
$c_0$, $c_2$, $c_4$, $v_0$, $v_2$) are chosen independently from a Gaussian 
distribution of random numbers with zero mean and width $\sigma$ 
\cite{BF1,BF2}. 

The connection between the IBM, its potential energy surfaces, equilibrium 
configurations and geometric shapes, can be studied with mean-field 
Hartree-Bose techniques by means of coherent states \cite{cs,duke}. 
Since the one- and two-body IBM Hamiltonian of Eq.~(\ref{ham}) does not 
give rise to a triaxial deformation \cite{PvI}, the coherent state can 
be written as an axially symmetric condensate 
\ba
\left| \, N,\alpha \, \right> \;=\; \frac{1}{\sqrt{N!}} 
\left( \cos \alpha \, s^{\dagger} + \sin \alpha \, d_0^{\dagger} 
\right)^N \, \left| \, 0 \, \right> ~, 
\ea
with $-\pi/2 < \alpha \leq \pi/2$. The angle $\alpha$ is related to 
the deformation parameters in the intrinsic frame, $\beta$ and $\gamma$ 
\cite{IBM,cs}. The potential energy surface is then 
given by the expectation value of the Hamiltonian in the coherent state 
\ba
E_N(\alpha) \;=\; a_4 \, \sin^4 \alpha + 
a_3 \, \sin^3 \alpha \cos \alpha + a_2 \, \sin^2 \alpha + a_0 ~, 
\ea
where the coefficients $a_i$ are linear combinations of the parameters 
of the Hamiltonian 
\ba
a_4 &=& \frac{1}{2} u_0 - u_2 
+ \frac{1}{10} c_0 + \frac{1}{7} c_2 + \frac{9}{35} c_4 
- \frac{1}{\sqrt{10}} v_0 ~, 
\nonumber\\
a_3 &=& -\sqrt{\frac{2}{7}} v_2 ~,
\nonumber\\
a_2 &=& \epsilon_d-\epsilon_s - u_0 + u_2 
+ \frac{1}{\sqrt{10}} v_0 ~, 
\nonumber\\
a_0 &=& \epsilon_s + \frac{1}{2} u_0 ~. 
\ea
For random interactions, the trial wave function 
$\left| \, N,\alpha \, \right>$ and the corresponding energy surface 
$E_N(\alpha)$ provide information on the distribution of shapes that the 
model can acquire. The equilibrium configuration, or the intrinsic 
vibrational state, is characterized by the 
value of $\alpha=\alpha_0$ for which the energy surface attains its 
minimum value (this procedure is equivalent to solving the Hartree-Bose 
equation). For a given Hamiltonian, the value of $\alpha_0$ depends 
on the coefficients $a_4$, $a_3$ and $a_2$. Hence, the distribution of 
shapes for an ensemble of Hamiltonians depends on the joint probability 
distribution $P(a_4,a_3,a_2)$, which for the present case is given by 
a multivariate normal distribution. In practice, for each Hamiltonian 
the minimum of the energy surface $E(\alpha)$ is determined numerically. 
The equilibrium configurations can be divided into three different classes: 
an $s$-boson or spherical condensate ($\alpha_0=0$), a deformed condensate 
with prolate or oblate symmetry ($0 < \alpha_0 < \pi/2$ or 
$-\pi/2 < \alpha_0 < 0$, respectively), and a $d$-boson condensate 
($\alpha_0=\pi/2$). 

\begin{figure}
\centerline{\hbox{\epsfig{figure=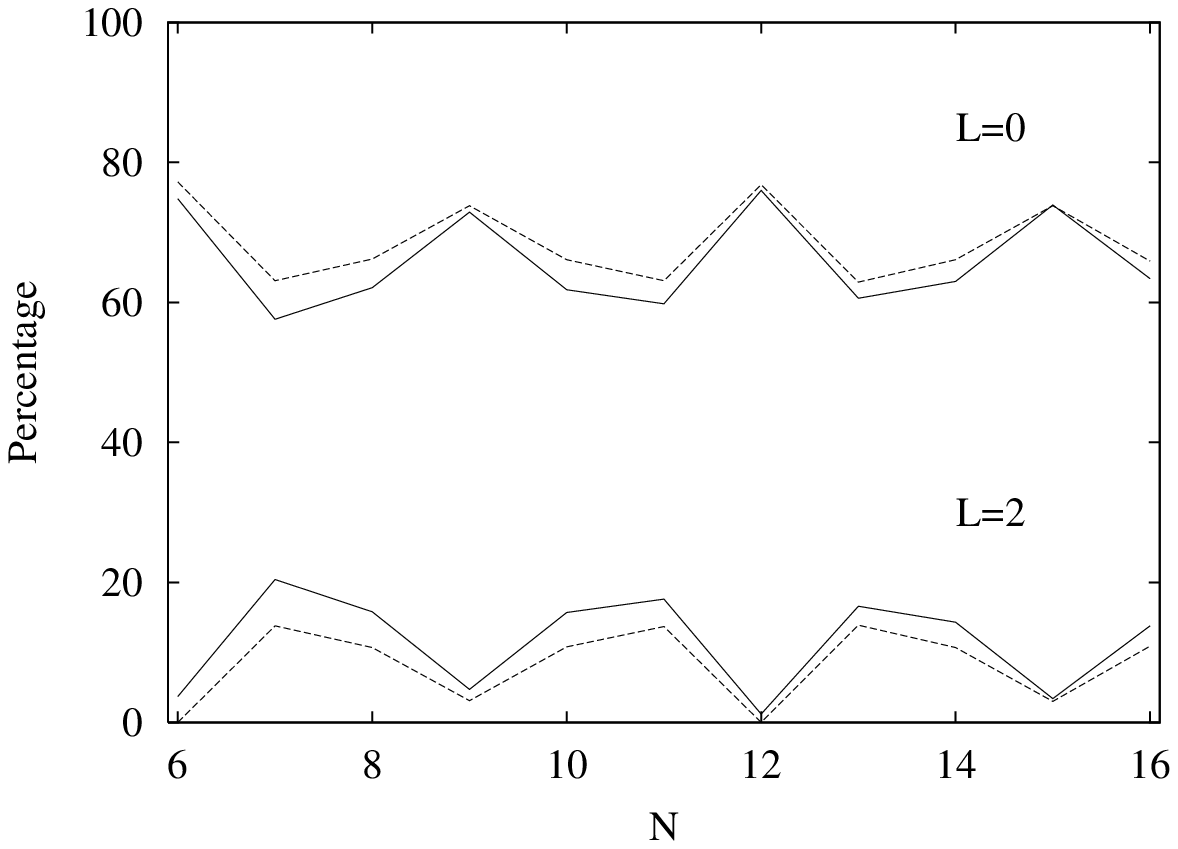,width=0.75\linewidth} }}
\caption[]{Percentages of ground states with $L=0$ and $L=2$ in the IBM 
with random one- and two-body interactions calculated exactly for 
10,000 runs (solid lines) and in mean-field approximation (dashed lines).}
\label{ibmgs}
\end{figure}

Each equilibrium configuration has its own characteristic angular momentum 
content. Even though we do not explicitly project the angular momentum 
states from the coherent state, the angular momentum of the ground state 
can be obtained from the rotational structure of the condensate in 
combination with the Thouless-Valatin formula for the corresponding moments 
of inertia. This procedure is described in detail in \cite{duke}. 
The results are summarized in Table~\ref{percibm}. 
\begin{itemize}

\item The $s$-boson condensate corresponds to a spherical shape. 
Whenever such a condensate occurs (in 39.4 $\%$ of the cases), 
the ground state has $L=0$.

\item The deformed condensate corresponds to an axially symmetric deformed 
rotor. The ordering of the rotational energy levels $L=0,2,\ldots,2N$ is 
determined by the sign of the moment of inertia 
\ba
E_{\rm rot} \;=\;  \frac{1}{2{\cal I}_3} L(L+1) ~. 
\ea
The deformed condensate occurs in 36.8 $\%$ of the cases. For ${\cal I}_3>0$ 
the ground state has $L=0$ (23.7 $\%$), while for ${\cal I}_3<0$ the ground 
state has the maximum value of the angular momentum $L=2N$ (13.1 $\%$). 

\item The $d$-boson condensate corresponds to a quadrupole oscillator with 
$N$ quanta. Its rotational structure has a more complicated structure 
than the previous two cases. It is characterized by the labels $\tau$, 
$n_{\Delta}$ and $L$. The boson seniority $\tau$ is given by 
$\tau=3n_{\Delta}+\lambda=N,N-2,\ldots,1$ or 
$0$ for $N$ odd or even, and the values of the angular momenta are 
$L=\lambda,\lambda+1,\ldots,2\lambda-2,2\lambda$ \cite{IBM}. In this case, 
the rotational excitation energies depend on two moments of inertia 
\ba
E_{\rm rot} \;=\; \frac{1}{2{\cal I}_5} \tau(\tau+3) 
+ \frac{1}{2{\cal I}_3} L(L+1) ~, 
\ea
which are associated with the spontaneously broken three- and 
five-dimensional rotational symmetries of the $d$-boson condensate. 
The $d$-boson condensate occurs in the remaining 23.8 $\%$ of the cases. 
The results in Table~\ref{percibm} can be understood qualitatively as 
follows. For ${\cal I}_5>0$ the ground state has $\tau=0$ 
for $N$ even or $\tau=1$ for $N$ odd ($\sim$ 4 $\%$), while for 
${\cal I}_5<0$ the ground state has the maximum value of the boson 
seniority $\tau=N$ ($\sim$ 19 $\%$). For $\tau=0$ and $\tau=1$ 
there is a single angular momentum state with $L=0$ and $L=2$, 
respectively. For  the $\tau=N$ multiplet, the angular momentum of the 
ground state depends on the sign of the moment of inertia ${\cal I}_3$. 
For ${\cal I}_3>0$ the ground state has $L=0$ for $N=3k$ or 
$L=2$ for $N \neq 3k$ (9 $\%$), while for ${\cal I}_3<0$ the ground 
state has the maximum value of the angular momentum $L=2N$ (10 $\%$). 

\end{itemize}
Table~\ref{percibm} shows that the spherical and deformed 
condensates contribute constant amounts of 39.4 $\%$ and 23.7 $\%$, 
respectively, to the $L=0$ ground state percentage, whereas the contribution 
from the $d$-boson condensate depends on the number of bosons $N$. The 
$L=2$ ground states arise completely from the $d$-boson condensate solution. 
In Fig.~\ref{ibmgs} we show the percentages of ground states with 
$L=0$ and $L=2$ as a function of the total number of bosons $N$. 
A comparison of the results of the mean-field analysis (dashed lines) 
and the exact ones (solid lines) shows a good agreement.  
There is a dominance of ground states with $L=0$ for $\sim$ 63-77 
$\%$ of the cases. Both for $L=0$ and $L=2$ there are large oscillations 
with $N$, which are entirely due to the contribution of the $d$-boson 
condensate. For $N=3k$ we see an enhancement for $L=0$ and a 
corresponding decrease for $L=2$. In the mean-field analysis, the sum of the 
two, which corresponds to $\sim$ 77 $\%$ of the cases, hardly depends on the 
number of bosons, in agreement with the exact results. 
For the remaining $\sim$ 23 $\%$ of the cases, 
the ground state has the maximum value of the angular momentum $L=2N$. 

\begin{figure}
\centerline{\hbox{\epsfig{figure=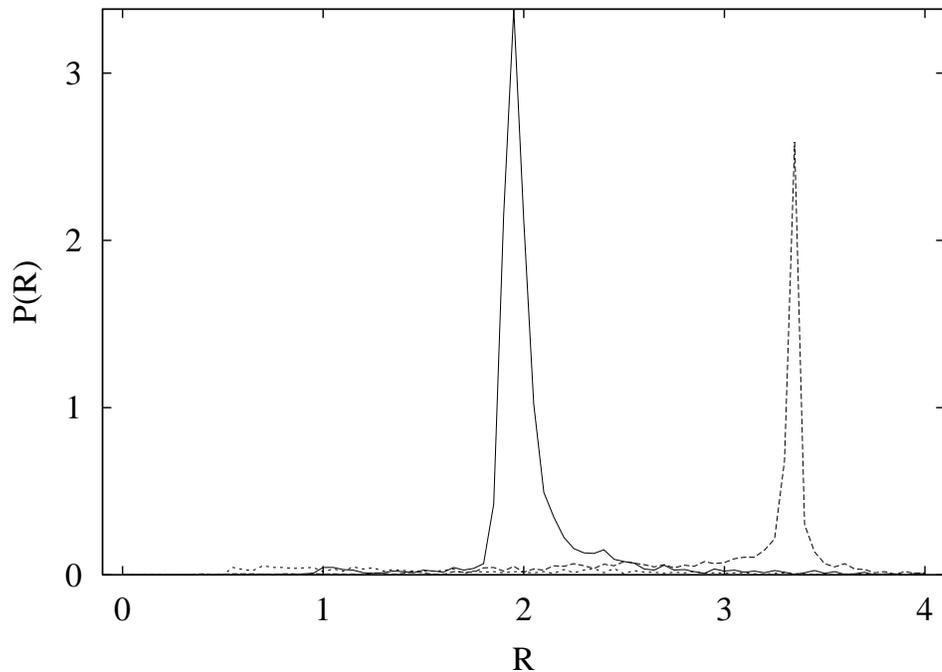,width=0.75\linewidth} }}
\caption[]{Probability distribution $P(R)$ of the energy ratio $R$ 
obtained for $N=16$ and 10,000 runs 
for the spherical (solid line), deformed (dashed line) and $d$-boson 
condensate (dotted line) equilibrium configurations, respectively.} 
\label{ratio16}
\end{figure}

\begin{figure}
\centerline{\hbox{\epsfig{figure=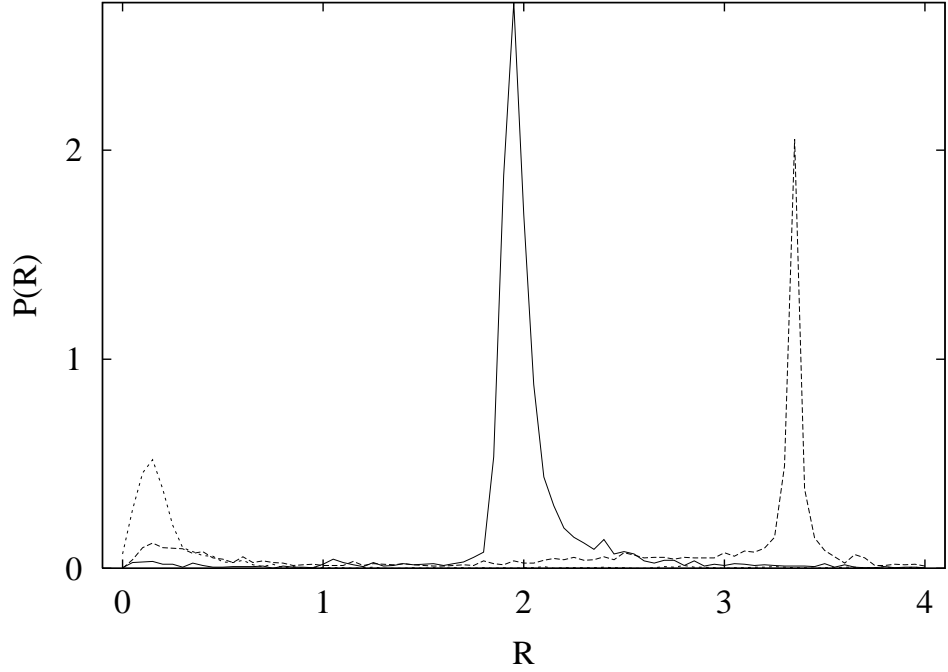,width=0.75\linewidth} }}
\caption[]{As Fig.~\protect\ref{ratio16}, but for $N=15$.} 
\label{ratio15}
\end{figure}

For the cases with an $L=0$ ground state, the probability distribution $P(R)$ 
of the energy ratio 
\ba
R \;=\; \frac{E_{4_1}-E_{0_1}}{E_{2_1}-E_{0_1}} ~,  
\ea
can be used to study the characteristics of the energy spectra. 
As mentioned above, in a 
numerical study of the IBM with random interactions \cite{BF1} it was found 
that $P(R)$ exhibits two very pronounced peaks, right at the vibrational 
value of $R=2$ and at the rotational value of $R=10/3$. This provides a 
clear indication of the occurrence of vibrational and rotational structure, 
which was confirmed by a simultaneous study of the quadrupole transitions 
between the levels \cite{BF1}. In Figs.~\ref{ratio16} and~\ref{ratio15} we 
show the contribution of each one of the equilibrium configurations to $P(R)$ 
for $N=16$ and $N=15$, respectively. For both cases, the spherical shape 
(solid line) contributes almost exclusively to the peak at $R=2$, and 
similarly the deformed shape (dashed line) to the peak at $R=10/3$, which 
once again confirms the vibrational and rotational character of these maxima. 
For $N=16$ the contribution of the $d$-boson condensate (dotted line) 
is small, whereas for $N=15$ it gives a contribution for small values of $R$, 
which corresponds to a level sequence $L=0$, $4$, $2$. 

\begin{figure}
\centerline{\hbox{\epsfig{figure=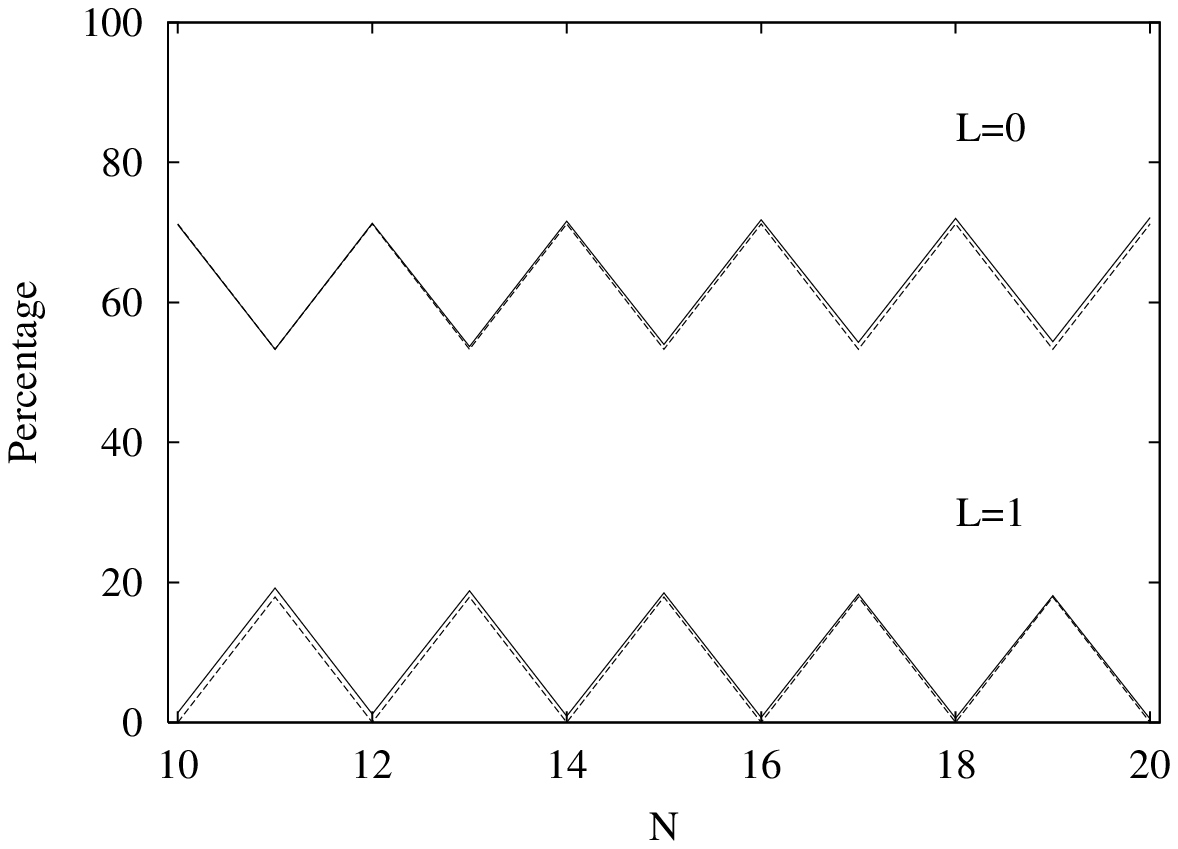,width=0.75\linewidth} }}
\caption[]{Percentages of ground states with $L=0$ and $L=1$ in the vibron 
model with random one- and two-body interactions calculated exactly for 
100,000 runs (solid lines) and in mean-field approximation (dashed lines).}
\label{vibgs}
\end{figure}

The above analysis is not only valid for the IBM, but can also be applied 
to other many-body systems of randomly interacting bosons. As an example, 
we mention the vibron model which was introduced to describe the 
relative motion in two-body problems, e.g. diatomic molecules \cite{vibron}, 
nuclear clusters \cite{cluster} and mesons \cite{meson}. The vibron model 
has many of the same qualitative features as the IBM, namely vibrational 
and rotational spectra, but has a much simpler mathematical structure. 
Its building blocks are a dipole boson $p^{\dagger}$ with $L^P=1^-$ and a 
scalar boson $s^{\dagger}$ with $L^P=0^+$. The total number of bosons $N$ 
is conserved by the Hamiltonian. In this case, the coherent state is 
expressed as a condensate of a superposition of a scalar and a dipole 
boson \cite{onno}
\ba
\left| \, N,\alpha \, \right> \;=\; \frac{1}{\sqrt{N!}} 
\left( \cos \alpha \, s^{\dagger} + \sin \alpha \, p_0^{\dagger} 
\right)^N \, \left| \, 0 \, \right>  ~, 
\ea
with $0 \leq \alpha \leq \pi/2$. The potential energy surface of a one- and 
two-body vibron Hamiltonian is a quadratic function of $\sin^2 \alpha$ 
\ba
E_N(\alpha) \;=\; a_4 \, \sin^4 \alpha + a_2 \, \sin^2 \alpha + a_0 ~. 
\ea
In this case, the value of $\alpha_0$ that characterizes the equilibrium 
configuration only depends on the two coefficients $a_4$ and $a_2$. 
The distribution of shapes for an ensemble of Hamiltonians is determined by 
the joint probability distribution $P(a_4,a_2)$. For the vibron model, 
the structure of the energy surface is simpler than for the IBM. This 
makes it possible to obtain analytic results for the probability of the 
occurrence of a given equilibrium configuration. As for the IBM, the 
solutions can be divided into three different classes. 
\begin{itemize}

\item The $s$-boson or spherical condensate ($\alpha_0=0$) occurs with 
probability 
\ba
P_1 \;=\; \frac{1}{4\pi} \left( \pi+2 \arctan \sqrt{\frac{27}{16}} 
\right) \;=\; 0.396 ~. 
\ea 
This solution corresponds to a spherical shape which has $L=0$. 

\item The deformed condensate ($0 < \alpha_0 < \pi/2$) has probability 
\ba
P_2 \;=\; \frac{1}{2\pi} \arctan \sqrt{\frac{64}{3}} \;=\; 0.216 ~.
\ea
The ordering of the rotational energy levels $L=0,1,\ldots,N$ is 
determined by the sign of the moment of inertia 
\ba
E_{\rm rot} \;=\;  \frac{1}{2{\cal I}_3} L(L+1) ~. 
\ea
For ${\cal I}_3>0$ the ground state has $L=0$ (13.8 $\%$), while for 
${\cal I}_3<0$ the ground state has the maximum value of the angular 
momentum $L=N$ (7.8 $\%$). 

\item The $p$-boson condensate ($\alpha_0=\pi/2$) has probability 
\ba
P_3 \;=\; 1-P_1-P_2 \;=\; 0.388 ~. 
\ea
This solutions corresponds to a three-dimensional harmonic oscillator with 
$N$ quanta. Its angular momentum content is given by $L=N,N-2,\ldots,1$ 
or $0$ for $N$ odd or even, respectively. 
For ${\cal I}_3>0$ the ground state has $L=0$ for $N$ even or $L=1$ for 
$N$ odd (17.9 $\%$), while for ${\cal I}_3<0$ the ground state has the 
maximum value of the angular momentum $L=N$ (20.9 $\%$).  

\end{itemize}
Table~\ref{percvib} shows that the spherical and deformed condensates 
contribute a constant amount of 53.4 $\,\%$ to the $L=0$ ground state 
percentage. For the $p$-boson condensate, the angular momentum of the ground 
state depends on the sign of the moment of inertia and the number of bosons 
$N$. For even values of $N$, the ground state either has $L=0$ or $L=N$,  
whereas for odd values it has either $L=1$ or $L=N$. As a result, 
the vibron model shows a dominance of $L=0$ ground states which oscillates
between $\sim$ 53 $\%$ for odd values of $N$ and $\sim$ 71 $\%$ for even 
values. The $L=1$ ground states arise completely from the $p$-boson 
condensate solution. 
In Fig.~\ref{vibgs} we show the percentages of ground states with 
$L=0$ and $L=1$ as a function of the total number of bosons $N$. 
A comparison of the results of the mean-field analysis (dashed lines) 
and the exact ones (solid lines) shows an excellent agreement.  
The oscillations in the percentages of $L=0$ and $L=1$ ground states are 
entirely due to the contribution of the $p$-boson condensate, whereas their 
sum is a constant which does not depend on $N$. 

In this paper, we have studied the properties of low-lying states in 
the IBM  with random interactions. We addressed the 
origin of the regular features, that were obtained before in numerical 
studies of the IBM, in particular the dominance of $L=0$ ground states and 
the occurrence of vibrational and rotational band structures. We have shown 
that a mean-field analysis can explain these features and provide a more 
intuitive understanding of their origin. 
The use of mean-field techniques and coherent states 
circumvents the use of coefficients of fractional parentage, bypasses 
the diagonalization of thousands of matrices, and makes it possible to 
associate different regions of the parameter space with particular 
intrinsic vibrational states, which in turn correspond to 
definite geometric shapes. 
For the IBM, there are three different equilibrium configurations or 
geometric shapes: a spherical shape ($\sim$ 39 $\%$), a deformed shape 
($\sim$ 36 $\%$) and a condensate of quadrupole bosons ($\sim$ 25 $\%$). 
Since the spherical shape only has $L=0$, and the deformed shape has this 
value of $L$ in about two thirds of the cases, these two solutions account 
for $\sim$ 63 $\%$ of $L=0$ ground states. The oscillations observed 
for the $L=0$ ground state percentage can be ascribed completely to 
the contribution of the $d$-boson condensate. In addition, 
we found a one-to-one correspondence between the occurrence of the 
spherical and deformed equilibrium configurations and the peaks in the 
probability distribution for the energy ratio $R=2$ and $R=10/3$, 
respectively. Similar conclusions hold for the vibron model. 

The study of interacting boson models with random interactions indicate 
that there is a significantly larger class of Hamiltonians that leads to 
regular, ordered behavior at the low excitation energies than was commonly 
assuemd. The fact that these properties are shared by both the 
IBM and the vibron model, seems to exclude an explanantion 
solely in terms of the angular momentum algebra, the connectivity of the 
model space, or the many-body dynamics of the model, as has been suggested 
before. The present analysis points to a more general phenomenon that does 
not depend so much on the details of the angular momentum coupling, but 
rather on the occurrence of definite, robust geometric 
phases such as spherical and deformed shapes. 

Since the collective subspace involves a drastic truncation of the shell 
model space as well as a bosonic approximation, our conclusions cannot be 
applied directly to the fermion space. We are currently exploring a 
nucleon-pair truncation scheme in order to search for answers in this more 
complex situation \cite{Bartol}. These results may help us understand the 
appearance of robust properties in many-body quantum systems with random 
interactions, be it in nuclear, atomic, molecular 
or mesoscopic systems like quandom dots \cite{yoram}. 

\section*{Acknowledgements}

This work was supported in part by CONACyT under projects 
32416-E and 32397-E, and by DPAGA-UNAM under project IN106400.

\clearpage

\begin{table}
\centering
\caption[]{Percentages of ground states with $L=0$, $2$ and $2N$, obtained 
in a mean-field analysis of the interacting boson model.}
\label{percibm}
\vspace{15pt}
\begin{tabular}{crrrl}
& & & & \\
Shape & $L=0$ & $L=2$ & $L=2N$ & \\
& & & & \\
\hline
& & & & \\
$  \alpha_0=0$   & 39.4 $\,\%$ &  0.0 $\,\%$ &  0.0 $\,\%$ & \\
& & & & \\
$0<|\alpha_0|<1$ & 23.7 $\,\%$ &  0.0 $\,\%$ & 13.1 $\,\%$ & \\
& & & & \\
$  \alpha_0=1$   & 13.5 $\,\%$ &  0.0 $\,\%$ & 10.3 $\,\%$ & $N=6k$ \\
&  0.2 $\,\%$ & 13.2 $\,\%$ & 10.4 $\,\%$ & $N=6k+1,6k+5$ \\
&  4.4 $\,\%$ &  9.0 $\,\%$ & 10.4 $\,\%$ & $N=6k+2,6k+4$ \\
&  9.3 $\,\%$ &  4.0 $\,\%$ & 10.5 $\,\%$ & $N=6k+3$ \\
& & & & \\
\end{tabular}
\end{table}

\begin{table}
\centering
\caption[]{Percentages of ground states with $L=0$, $1$ and $N$, obtained 
in a mean-field analysis of the vibron model.}
\label{percvib}
\vspace{15pt}
\begin{tabular}{crrrl}
& & & & \\
Shape & $L=0$ & $L=1$ & $L=N$ & \\
& & & & \\
\hline
& & & & \\
$  \alpha_0=0$ & 39.6 $\,\%$ &  0.0 $\,\%$ &  0.0 $\,\%$ & \\
& & & & \\
$0<\alpha_0<1$ & 13.8 $\,\%$ &  0.0 $\,\%$ &  7.8 $\,\%$ & \\
& & & & \\
$  \alpha_0=1$ & 17.9 $\,\%$ &  0.0 $\,\%$ & 20.9 $\,\%$ & $N=2k$ \\
               &  0.0 $\,\%$ & 17.9 $\,\%$ & 20.9 $\,\%$ & $N=2k+1$ \\
& & & & \\
\end{tabular}
\end{table}

\clearpage

\clearpage

\begin{thebibliography}{aa}

\bibitem{Zamfir}
N.V. Zamfir, R.F. Casten and D.S. Brenner,
Phys. Rev. Lett. {\bf 72}, 3480 (1994).

\bibitem{JBD}
C.W. Johnson, G.F. Bertsch and D.J. Dean,
Phys. Rev. Lett. {\bf 80}, 2749 (1998).

\bibitem{BFP1}
R. Bijker, A. Frank and S. Pittel,
Phys. Rev. C {\bf 60}, 021302 (1999).

\bibitem{JBDT}
C.W. Johnson, G.F. Bertsch, D.J. Dean and I. Talmi,
Phys. Rev. C {\bf 61}, 014311 (2000).

\bibitem{BF1}
R. Bijker and  A. Frank,
Phys. Rev. Lett. {\bf 84}, 420 (2000).

\bibitem{BF2} 
R. Bijker and A. Frank, 
Phys. Rev. C {\bf 62}, 014303 (2000).
 
\bibitem{BFP2}
R. Bijker, A. Frank and S. Pittel, 
Rev. Mex. F{\'{\i}}s. {\bf 46 S1}, 47 (2000). 

\bibitem{KZC}
D. Kusnezov, N.V. Zamfir and R.F. Casten, 
Phys. Rev. Lett. {\bf 85}, 1396 (2000). 

\bibitem{MVZ} 
D. Mulhall, A. Voyla and V. Zelevinsky, 
Phys. Rev. Lett. {\bf 85}, 4016 (2000). 

\bibitem{DK}
D. Kusnezov, 
Phys. Rev. Lett. {\bf 85}, 3773 (2000); 
R. Bijker and A. Frank, 
Phys. Rev. Lett {\bf 87}, 029201 (2001);   
D. Kusnezov, 
Phys. Rev. Lett {\bf 87}, 029202 (2001).  

\bibitem{Kaplan}
L. Kaplan, T. Papenbrock and C.W. Johnson, 
Phys. Rev. {\bf 63}, 4307 (2001). 

\bibitem{Drozdz}
S. Dro\v{z}d\v{z} and M. W\'ojcik, 
preprint nucl-th/0007045.

\bibitem{ZA}
Y.M. Zhao and A. Arima, 
preprint nucl-th/0108052. 

\bibitem{victor}
V. Vel\'azquez and A.P. Zuker, 
preprint nucl-th/0106020. 

\bibitem{IBM}
F. Iachello and A. Arima,
{\it The interacting boson model}
(Cambridge University Press, 1987). 

\bibitem{cs}
J.N. Ginocchio and M. Kirson, 
Phys. Rev. Lett. {\bf 44}, 1744 (1980); 
A.E.L. Dieperink, O. Scholten and F. Iachello, 
Phys. Rev. Lett. {\bf 44}, 1747 (1980). 

\bibitem{duke}
J. Dukelsky, G.G. Dussel, R.P.J. Perazzo, S.L. Reich and H.M. Sofia, 
Nucl. Phys. A {\bf 425}, 93 (1984). 

\bibitem{PvI}
P. van Isacker and J.Q. Chen, 
Phys. Rev. C {\bf 24}, 684 (1981). 

\bibitem{vibron}
F. Iachello, 
Chem. Phys. Lett. {\bf 78}, 581 (1981).  

\bibitem{cluster}
F. Iachello, 
Phys. Rev. C {\bf 23}, 2778 (1981).

\bibitem{meson}
F. Iachello, N.C. Mukhopadhyay and L. Zhang, 
Phys. Rev. D {\bf 44}, 898 (1991). 

\bibitem{onno}
O.S. van Roosmalen and A.E.L. Dieperink, 
Ann. Phys. (N.Y.) {\bf 139}, 198 (1982); 
S. Levit and U. Smilansky, 
Nucl. Phys. A {\bf 389}, 56 (1982). 

\bibitem{Bartol}
Y.M. Zhao, S. Pittel, R. Bijker, A. Frank and A. Arima, 
to be published. 

\bibitem{yoram}
Y. Alhassid, 
Rev. Mod. Phys. {\bf 72}, 895 (2000).

\end{thebibliography}
\end{document}